\documentclass[aps,showpacs,onecolumn,11pt]{revtex4}
\usepackage{epsfig,colordvi}
\usepackage{graphics}
\usepackage{amssymb,amsmath}
\usepackage{times}
\newcommand{\ba}{\begin{eqnarray}}
\newcommand{\ea}{\end{eqnarray}}
\newcommand{\bsub}{\begin{subequations}}
\newcommand{\esub}{\end{subequations}}

\def\ket#1{|#1\rangle}
	
\def\bsu3{\overline{{\rm SU(3)}}}
\def\bso6{\overline{{\rm SO(6)}}}
\def\bPi2{\overline{\Pi}^{(2)}}
\def\blam{\bar{\lambda}}
\def\bmu{\bar{\mu}}
\def\bK{\bar{K}}
\def\b0{\beta_0}
\def\beq{\beta_{\rm eq}}
\def\g0{\gamma_0}
\def\gaeq{\gamma_{\rm eq}}

\begin{document}

\sloppy \raggedbottom

\setcounter{page}{1}

\title{Symmetry-Based Approach to Shape~Coexistence in Nuclei}

\author{A. Leviatan and N. Gavrielov}{}

\address{Racah Institute of Physics, The Hebrew University,
Jerusalem 91904, Israel}{}

%\received{20 December 2017}

\begin{abstract}
A symmetry-based approach for describing shape-coexistence, 
is presented in the framework of the interacting boson model of nuclei.  
It involves a construction of a number-conserving Hamiltonian 
which preserves the dynamical symmetry of selected bands 
associated with each shape, while breaking the symmetries in other states.
The resulting structure embodies multiple partial dynamical symmetries. 
The procedure is applied to prolate-oblate and spherical-prolate-oblate 
coexistence, at and slightly away from the critical points.
\end{abstract}
\pacs{21.60.Fw, 21.10.Re, 21.60.Ev}
\maketitle

\section[]{Introduction}

The presence of distinct shapes at similar low energies 
in a given nucleus, is a phenomena known to occur widely 
across the nuclear chart~\cite{Heyde11}, 
including nuclei far from stability~\cite{Jenkins14}.
Notable empirical examples include the coexistence of prolate and 
oblate shapes in Kr~\cite{Clement07}, 
Se~\cite{Ljun08} and Hg isotopes~\cite{Bree14}, 
and the triple coexistence of spherical, prolate and oblate shapes 
in $^{186}$Pb~\cite{Andreyev00}. 
A~detailed microscopic interpretation of nuclear shape-coexistence
is a formidable task. In a shell model description of nuclei 
near shell-closure, it is attributed to the occurrence of 
multi-particle multi-hole intruder excitations across shell gaps. 
For medium-heavy nuclei, this necessitates drastic truncations of 
large model spaces, {\it e.g.}, by a bosonic approximation of nucleon 
pairs~\cite{DuvBar81,Foisson03,Frank04,Morales08,ramos14,
nomura13,Nomura16}. 
In a mean-field approach, based on energy density functionals, 
the coexisting shapes are associated with different minima 
of an energy surface calculated self-consistently. 
A detailed comparison with spectroscopic observables 
requires beyond mean-field methods, 
including restoration of broken symmetries and configuration mixing of 
angular-momentum and particle-number projected 
states~\cite{Bender13,vretenar16}. 
Such extensions present a major computational effort and often 
require simplifying assumptions such as a mapping to collective model 
Hamiltonians~\cite{EDF-IBM11}.
In the present contribution, we consider a simple 
alternative to describe shape coexistence, 
in the framework of the interacting boson model~(IBM)~\cite{ibm} of nuclei. 
The proposed approach is founded on the notion of partial dynamical symmetry 
(PDS)~\cite{Leviatan11}, emphasizing the role of remaining underlying 
symmetries which provide physical insight and make the problem tractable.

\section[]{Dynamical Symmetries and Nuclear Shapes in the IBM}

The IBM has been widely used to describe quadrupole
collective states in nuclei
in terms of $N$ monopole ($s^\dag$) and quadrupole ($d^\dag$) bosons,
representing valence nucleon pairs. 
The model has U(6) as a spectrum generating algebra 
and its solvable limits correspond to dynamical symmetries 
associated with the following chains of nested sub-algebras of U(6)
\bsub
\ba
&&{\rm U(6)\supset U(5)\supset SO(5)\supset SO(3)} \;\;\,\quad
\quad\ket{N,\, n_d,\,\tau,\,n_{\Delta},\,L} ~,\quad 
\label{U5-ds}
\\
&&{\rm U(6)\supset SU(3)\supset SO(3)} \;\;\;\;\qquad\quad\quad
\quad\ket{N,\, (\lambda,\mu),\,K,\, L}~,\quad 
\label{SU3-ds}
\\
&&{\rm U(6)\supset \bsu3\supset SO(3)} \;\;\;\;\qquad\quad\quad
\quad\ket{N,\, (\blam,\bmu),\,\bar{K},\, L} ~,\quad
\label{SU3bar-ds}
\\
&&
{\rm U(6)\supset SO(6)\supset SO(5)\supset SO(3)} \,\quad
\quad\ket{N,\, \sigma,\,\tau,\,n_{\Delta},\, L} ~.\quad 
\label{O6-ds}
\ea
\label{IBMchains}
\esub
A dynamical symmetry~(DS) occurs when the Hamiltonian is expressed 
in terms of the Casimir operators of a given chain, in which case, 
all states are solvable and classified by
the indicated quantum numbers which are the labels of 
irreducible representations (irreps) of the algebras in the chain.
The analytic solutions corresponding to the above DS chains, 
with leading subalgebras: U(5), SU(3), ${\rm\overline{SU(3)}}$ and SO(6), 
resemble known paradigms of nuclear collective 
structure: spherical vibrator, prolate-, oblate- and $\gamma$-soft deformed 
rotors, respectively.  
This identification is consistent with the geometric visualization 
of the model, obtained by an energy surface, $E_{N}(\beta,\gamma)$, 
defined by the expectation value of the Hamiltonian in the coherent 
(intrinsic) state~\cite{gino80,diep80},
\bsub
\ba
\vert\beta,\gamma ; N \rangle &=&
(N!)^{-1/2}(b^{\dagger}_{c})^N\,\vert 0\,\rangle ~,
\\[1mm]
b^{\dagger}_{c} &=& (1+\beta^2)^{-1/2}[\beta\cos\gamma 
d^{\dagger}_{0} + \beta\sin{\gamma} 
( d^{\dagger}_{2} + d^{\dagger}_{-2})/\sqrt{2} + s^{\dagger}] ~.\qquad
\ea
\label{int-state}
\esub
Here $(\beta,\gamma)$ are
quadrupole shape parameters whose values, $(\beta_{\rm eq},\gamma_{\rm eq})$, 
at the global minimum of $E_{N}(\beta,\gamma)$ define the equilibrium 
shape for a given Hamiltonian.
The equilibrium deformations associated with the 
DS limits conform with their geometric interpretation 
and are given by
\bsub
\ba
&&{\rm U(5):} \qquad\;\;\, \beta_{\rm eq}=0 \hspace{2.1cm}
\quad\;\, n_d=0~,\quad\\
&&{\rm SU(3):} \qquad (\beta_{\rm eq} \!=\!\sqrt{2},\gamma_{\rm eq}\!=\!0)
\qquad\;\;\; (\lambda,\mu)=(2N,0)~,\quad\\
&&{\bsu3:} \qquad (\beta_{\rm eq} \!=\!\sqrt{2},\gamma_{\rm eq}\!=\!\pi/3)
\qquad (\blam,\bmu)=(0,2N)~,\quad\\
&& {\rm SO(6):} \qquad (\beta_{\rm eq}\!=\!1,\gamma_{\rm eq}\,\,{\rm arbitrary})
\quad\; \sigma =N ~. 
\ea
\label{int-states-ds}
\esub
For these values, as shown, the equilibrium intrinsic state 
$\ket{\beq,\gaeq;N}$ 
representing the ground band, becomes a lowest (or highest) weight state 
in a particular irrep of the leading sub-algebra 
in each of the chains of Eq.~(\ref{IBMchains}). 
The DS Hamiltonians support a single minimum in their 
energy surface, hence serve as benchmarks for the dynamics of 
a single quadrupole shape. 

\section[]{Partial Dynamical Symmetries and Shape Coexistence}

A dynamical symmetry (DS) is characterized by {\it complete} solvability 
and good quantum numbers for {\it all} states. Often the symmetry 
in question is obeyed by only selected states, {\it e.g.} members of the 
ground band in deformed nuclei. 
The need to address such situations, but still preserve 
important symmetry remnants, has lead to the introduction of 
partial dynamical symmetry (PDS)~\cite{Leviatan11,Alhassid92}. 
The latter is a generalization of the DS concept, and corresponds 
to a particular symmetry breaking for which 
only {\it some} of the states retain solvability 
and/or have good quantum numbers. 
In the present contribution, we show that this novel symmetry notion 
can play a vital role in formulating algebraic benchmarks for 
the dynamics of multiple quadrupole shapes.
We focus on the dynamics in the vicinity of the critical 
point, where the corresponding multiple minima in the energy surface 
are near-degenerate and the structure changes most rapidly.

Consider one of the DS chains of Eq.~(\ref{IBMchains}), 
\ba
{\rm U(6)\supset G_1\supset G_2\supset \ldots \supset SO(3)} &&\;\;\quad
\ket{N,\, \lambda_1,\,\lambda_2,\,\ldots,\,L} ~,\quad 
\label{u6-ds}
\ea                        
with leading sub-algebra $G_1$, related basis and associated shape 
$(\beta_{\rm eq},\gamma_{\rm eq})$. 
The construction of an Hamiltonian with PDS is 
done by requiring it to satisfy
\ba
\hat{H}\ket{\beq,\gaeq;N,\lambda_1\!=\!\Lambda_0,\lambda_2,\ldots,L} = 0 ~.
\label{Hvanish}
\ea 
The set of zero-energy eigenstates in Eq.~(\ref{Hvanish}) 
are basis states of a particular $G_1$-irrep, $\lambda_1=\Lambda_0$, 
and have good $G_1$ symmetry.
For a positive-definite $\hat{H}$, they span the ground band of 
the equilibrium shape and can be obtained by $L$-projection from 
the corresponding intrinsic state, 
$\vert\beta_{\rm eq},\gamma_{\rm eq} ; N\rangle$ of Eq.~(\ref{int-state}).
$\hat{H}$ itself, however, need not be invariant 
under $G_1$ and, therefore, has partial-$G_1$ symmetry. 
The Hamiltonian of Eq.~(\ref{Hvanish}) serves as the intrinsic part of the 
complete Hamiltonian, $\hat{H}'=\hat{H}+\hat{H}_c$.
Identifying the collective part ($\hat{H}_c$) with~the 
Casimir operators of 
the remaining sub-algebras of $G_1$ in the chain~(\ref{u6-ds}), 
the degeneracy of the above set of states 
is lifted, and they remain solvable eigenstates of $\hat{H}'$. 
The latter, by definition, has $G_1$-PDS and exemplifies 
an intrinsic-collective 
resolution~\cite{kirlev85,Leviatan87,levkir90,Leviatan06}, 
where the intrinsic part ($\hat{H}$) determines the energy 
surface, and the collective part ($\hat{H}_c$) is composed of kinetic 
rotational terms. IBM Hamiltonians with a single PDS, constructed 
in this manner, have been shown to be relevant to 
a broad range of nuclei with a single quadrupole 
shape~\cite{Leviatan96,LevSin99,Casten14,Couture15,
GarciaRamos09,Leviatan13,Kremer14,LevGino00,Pds-BF15}.

Coexistence of distinct shapes in the same nucleus, 
arises from competing terms in the Hamiltonian whose energy surface 
exhibits multiple minima, with different types of dynamics (and symmetry) 
associated with each minimum. In such circumstances, exact DSs are broken, 
and remaining symmetries, if any, are at most partial. 
A symmetry-based approach thus requires an extension of the above 
procedure to encompass a construction of Hamiltonians with several 
distinct PDSs~\cite{Leviatan07,Macek14,LevDek16,LevGav17}. 
For that purpose, consider two different shapes 
specified by equilibrium deformations 
($\beta_1,\gamma_1$) and ($\beta_2,\gamma_2$) whose dynamics is 
described, respectively, by the following DS chains
\bsub
\ba
{\rm U(6)\supset G_1\supset G_2\supset \ldots \supset SO(3)} &&\;\;\quad
\ket{N,\, \lambda_1,\,\lambda_2,\,\ldots,\,L} ~,\quad 
\label{ds-G1}\\
{\rm U(6)\supset G'_1\supset G'_2\supset \ldots \supset SO(3)} &&\;\;\quad
\ket{N,\, \sigma_1,\,\sigma_2,\,\ldots,\,L} ~,\quad 
\label{ds-G1prime}
\ea
\esub
with different leading sub-algebras ($G_1\neq G'_1$) and associated bases.
At the critical point, the corresponding minima representing the two shapes 
and the respective ground bands are degenerate. Accordingly, 
we require the intrinsic critical-point Hamiltonian to satisfy 
simultaneously the following two conditions
\bsub
\ba
\hat{H}\ket{\beta_1,\gamma_1;N,\lambda_1\!=\Lambda_0,\lambda_2,\ldots,L} 
&=& 0 ~,
\label{basis1}\\
\hat{H}\ket{\beta_2,\gamma_2;N,\sigma_1=\Sigma_0,\sigma_2,\ldots,L} 
&=&0 ~.
\label{basis2}
\ea
\label{bases12}
\esub
The states of Eq.~(\ref{basis1}) reside in the $\lambda_1=\Lambda_0$ irrep 
of $G_1$, are classified according to the DS-chain (\ref{ds-G1}), hence 
have good $G_1$ symmetry. Similarly,  
the states of Eq.~(\ref{basis2}) reside in the $\sigma_1=\Sigma_0$ irrep 
of $G'_1$, are classified according to the DS-chain (\ref{ds-G1prime}), 
hence have good $G'_1$ symmetry. Although $G_1$ and $G'_1$ are 
incompatible (non-commuting) symmetries, 
both sets are eigenstates of the same Hamiltonian. When the latter 
is positive definite, the two sets span the ground bands of the 
$(\beta_1,\gamma_1)$ and $(\beta_2,\gamma_2)$ shapes, respectively.
In general, $\hat{H}$ itself is not necessarily 
invariant under $G_1$ nor under $G_2$ and, therefore, its other eigenstates 
can be mixed under both $G_1$ and $G'_1$. 
Identifying the collective part of the Hamiltonian with the Casimir 
operator of SO(3) (as well as with the Casimir operators of additional 
algebras which are common to both chains), the two sets of states 
remain (non-degenerate) eigenstates of the complete Hamiltonian which 
then has both $G_1$-PDS and $G'_1$-PDS. 
The case of triple (or multiple) 
shape coexistence, associated with three (or more) incompatible DS-chains is 
treated in a similar fashion. 

The solution of Eqs.~(\ref{bases12}), if exists, results in a single 
number-conserving, rotational-invariant Hamiltonian with, possibly, 
higher-order terms. The effective Hamiltonian constructed in this manner, 
conserves the multiple DSs but only in selected bands.
This strategy is different from that used 
in the IBM with configuration 
mixing~\cite{DuvBar81,Foisson03,Frank04,Morales08}, 
where shape coexistence is described 
by different Hamiltonians for the normal 
and intruder configurations and a number-non-conserving mixing term.
In what follows, we apply the above procedure to a case study of double- 
and triple coexistence of prolate-oblate and 
spherical-prolate-oblate shapes.

\section{Prolate-Oblate and Spherical-Prolate-Oblate Shape Coexistence}

The DS limits appropriate to prolate and oblate shapes correspond 
to the chains~(\ref{SU3-ds}) and (\ref{SU3bar-ds}), respectively. 
For a given U(6) irrep $N$, the allowed SU(3) [$\,\bsu3\,$] irreps are 
$(\lambda,\mu)\!=\!(2N \!-\! 4k \!-\! 6m,2k)$ 
[$(\blam,\bmu)\!=\!(2k,2N\!-\!4k\!-\!6m)$] 
with $k,m$, non-negative integers. 
The multiplicity label $K$ ($\bK$) corresponds geometrically to the
projection of the angular momentum ($L$) on the symmetry axis. 
The basis states are eigenstates of the Casimir operator 
$\hat{C}_{2}[{\rm SU(3)}]$ or $\hat{C}_{2}[\bsu3]$, 
where $\hat{C}_{k}[G]$ denotes the Casimir operator of $G$ of order $k$.
Specifically, $\hat{C}_{2}[{\rm SU(3)}] \!=\! 2Q^{(2)}\cdot Q^{(2)} \!+\! 
{\textstyle\frac{3}{4}}L^{(1)}\cdot L^{(1)}$, 
$Q^{(2)} \!=\! d^{\dagger}s \!+\! s^{\dagger}\tilde{d} 
\!-\!\frac{1}{2}\sqrt{7} (d^{\dagger}\tilde{d})^{(2)}$, 
$L^{(1)} \!=\! \sqrt{10} (d^{\dagger}\tilde{d})^{(1)}$, 
$\tilde{d}_{\mu} \!=\! (-1)^{\mu}d_{-\mu}$
and $\hat{C}_{2}[\bsu3]$ is obtained by replacing $Q^{(2)}$ by 
$\bar{Q}^{(2)} \!=\! d^{\dagger}s \!+\! s^{\dagger}\tilde{d} 
\!+\!\frac{1}{2}\sqrt{7} (d^{\dagger}\tilde{d})^{(2)}$. 
The generators of SU(3) and $\bsu3$, $Q^{(2)}$ and $\bar{Q}^{(2)}$, 
and corresponding basis states, are related 
by a change of phase $(s^{\dag},s)\rightarrow (-s^{\dag},-s)$, 
induced by the operator ${\cal R}_s=\exp(i\pi\hat{n}_s)$, 
with $\hat{n}_s=s^{\dag}s$. 
The DS Hamiltonian involves a linear combination of the  Casimir operators 
in a given chain. The spectrum resembles that of an axially-deformed 
rotovibrator composed of SU(3) [or $\bsu3$] multiplets forming 
rotational bands, with $L(L+1)$-splitting generated by 
$\hat{C}_{2}[{\rm SO(3)}] \!=\! L^{(1)}\cdot L^{(1)}$. 
In the SU(3) [or $\bsu3$] DS limit, the lowest irrep $(2N,0)$ [or $(0,2N)$] 
contains the ground band $g(K\!=\!0)$ [or $g(\bK\!=\!0)$] 
of a prolate [oblate] deformed nucleus. 
The first excited irrep $(2N\!-\!4,2)$ [or $(2,2N\!-\!4)$] contains 
both the $\beta(K\!=\!0)$ and $\gamma(K\!=\!2)$ 
[or $\beta(\bK\!=\!0)$ and $\gamma(\bK\!=\!2)$] bands. 
Henceforth, we denote such prolate and oblate bands by 
$(g_1,\beta_1,\gamma_1)$ and ($g_2,\beta_2,\gamma_2$), respectively. 
Since ${\cal R}_sQ^{(2)}{\cal R}_s^{-1} \!=\! -\bar{Q}^{(2)}$, 
the SU(3) and $\bsu3$ DS spectra are identical and 
the quadrupole moments of corresponding states differ in sign. 

The U(5)-DS limit of Eq.~(\ref{U5-ds}) 
is appropriate to the dynamics of a spherical shape. 
For a given $N$, the allowed U(5) and SO(5) irreps 
are $n_d\!=\!0,1,2,\ldots, N$ and  $\tau\!=\!n_d,\,n_d\!-\!2,\dots 0$ 
or~$1$, respectively. 
The U(5)-DS spectrum resembles that of an anharmonic spherical vibrator, 
composed of U(5) $n_d$-multiplets whose spacing is governed by 
$\hat{C}_{1}[{\rm U(5)}]\!=\!\hat{n}_d=\sum_{\mu}d^{\dag}_{\mu}d_{\mu}$, and 
splitting is generated by the SO(5) and SO(3) terms. 
The lowest U(5) multiplets involve the ground state 
with quantum numbers $(n_d\!=\!0,\,\tau\!=\!0,\, L\!=\!0)$ 
and excited states with quantum numbers 
$(n_d=\!1\!,\,\tau\!=\!1,\, L\!=\!2)$ 
and $(n_d\!=\!2:\,\tau\!=\!0,\,L\!=\!0;\,\tau\!=\!2,\,L\!=\!2,4)$.

\begin{figure*}[t!]
  \centering
\includegraphics[width=5.5cm]{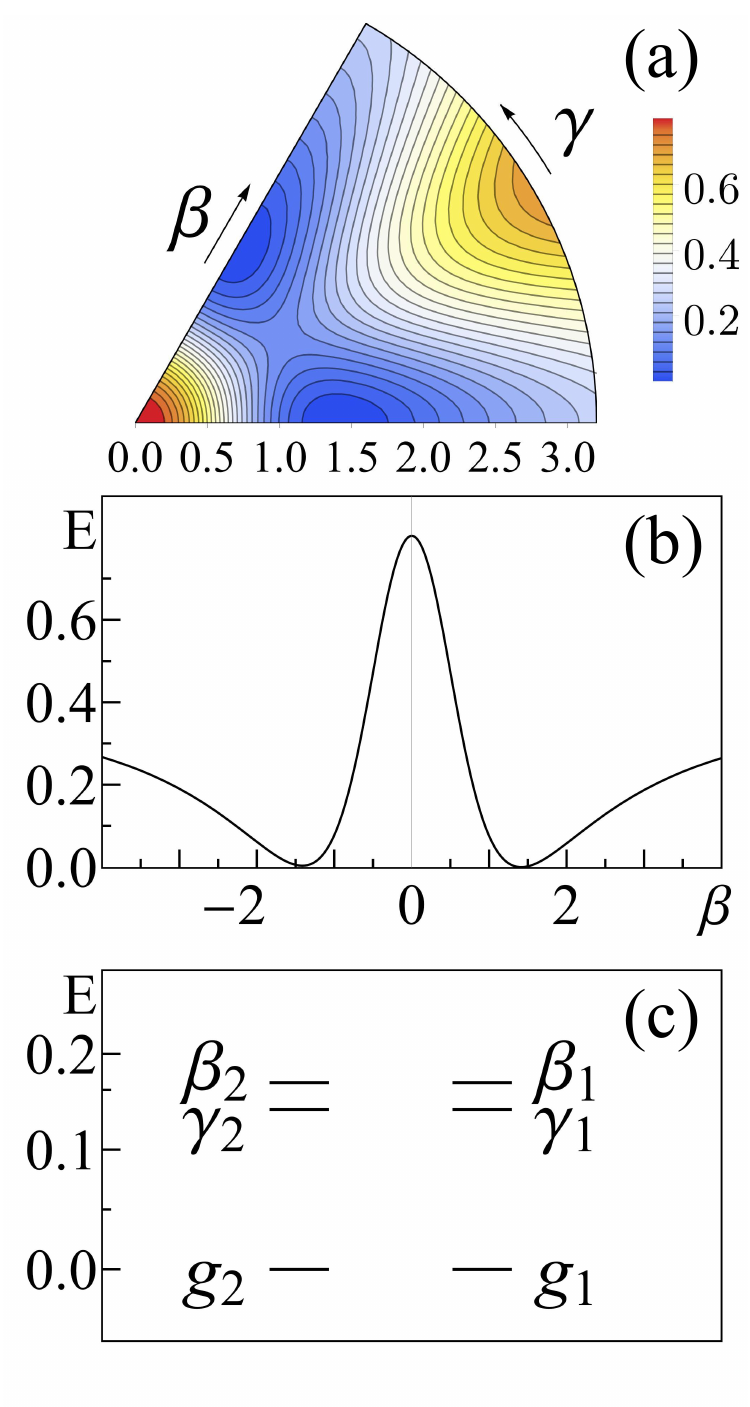}
\includegraphics[width=5.5cm]{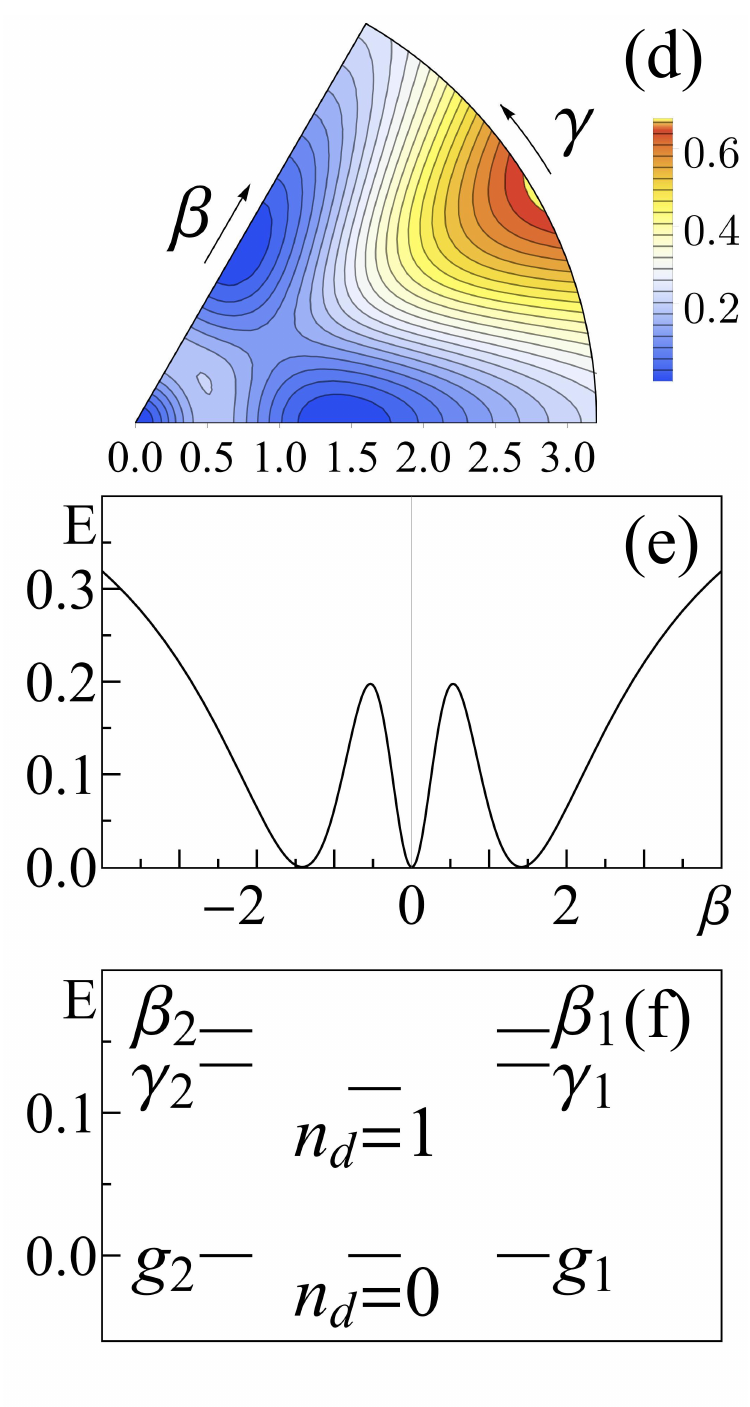}
  \caption{(Color online).
Contour plots of the energy surface~(\ref{surface}) [top row], 
$\gamma\!=\!0$ sections [middle row] and bandhead spectrum [bottom row] 
for the Hamiltonian~(\ref{Hprime}), with 
$\alpha\!=\!0.018,\,\eta_3=0.571,\,\rho\!=\!0,\,N\!=\!20$. 
Panels (a)-(b)-(c) [(d)-(e)-(f)] correspond to the choice 
$h_0\!=\!0.2,\,h_2\!=\!0.4$ [$h_0\!=\!0,\,h_2\!=\!0.5$] resulting in 
prolate-oblate [spherical-prolate-oblate] shape coexistence.
\label{fig1}}
\end{figure*}

Following the procedure of Eq.~(\ref{bases12}), 
the intrinsic part of the critical-point Hamiltonian, 
relevant to prolate-oblate (P-O) coexistence, is required to~satisfy
\bsub
\ba
\hat{H}\ket{N,\, (\lambda,\mu)=(2N,0),\,K=0,\, L} &=& 0 ~,
\label{2N0p}
\\
\hat{H}\ket{N,\, (\blam,\bmu)=(0,2N),\,\bar{K}=0,\, L} &=& 0 ~.
\label{02N}
\ea
\label{HvanishPO}
\esub 
Equivalently, $\hat{H}$ annihilates the intrinsic states of 
Eq.~(\ref{int-state}), with $(\beta\!=\!\sqrt{2},\gamma\!=\!0)$ and 
$(\beta\!=\!-\sqrt{2},\gamma\!=\!0)$, which are the lowest- and 
highest-weight vectors in the irreps $(2N,0)$ and $(0,2N)$ 
of SU(3) and $\bsu3$, respectively. 
The resulting Hamiltonian is found to be~\cite{LevDek16},
\ba
\hat{H} = 
h_0\,P^{\dag}_0\hat{n}_sP_0 + h_2\,P^{\dag}_0\hat{n}_dP_0 
+\eta_3\,G^{\dag}_3\cdot\tilde{G}_3 ~,
\label{Hint}
\ea
where $P^{\dag}_{0} = d^{\dagger}\cdot d^{\dagger} - 2(s^{\dagger})^2$,
$G^{\dag}_{3,\mu} = \sqrt{7}[(d^{\dag} d^{\dag})^{(2)}d^{\dag}]^{(3)}_{\mu}$, 
$\tilde{G}_{3,\mu} = (-1)^{\mu}G_{3,-\mu}$ 
and the centered dot denotes a scalar product. 
The corresponding energy surface, 
$E_{N}(\beta,\gamma) = N(N-1)(N-2)\tilde{E}(\beta,\gamma)$, 
is given by
\ba
\tilde{E}(\beta,\gamma) = 
\left\{(\beta^2-2)^2
\left [h_0 + h_2\beta^2\right ] 
+\eta_3 \beta^6\sin^2(3\gamma)\right \}(1+\beta^2)^{-3} ~.
\label{surface}
\ea
The surface is an even function of $\beta$ and 
$\Gamma = \cos 3\gamma$, 
and can be transcribed as 
$\tilde{E}(\beta,\gamma) = z_0 + 
(1+\beta^2)^{-3}[A\beta^6+ B\beta^6\Gamma^2 + D\beta^4+ F\beta^2]$, 
with $A \!=\! -4h_0 \!+\!h_2 \!+\! \eta_3,\, B \!=\! -\eta_3, \,
D \!=\! -(11h_0 \!+\! 4h_2), \; F \!=\! 4(h_2\!-\!4h_0),\,z_0 \!=\! 4h_0$. 
For $h_0,h_2,\eta_3\geq 0$, 
$\hat{H}$ is positive definite and 
$\tilde{E}(\beta,\gamma)$ has two degenerate global minima, 
$(\beta\!=\!\sqrt{2},\gamma\!=\!0)$ and 
$(\beta\!=\!\sqrt{2},\gamma\!=\!\pi/3)$ 
[or equivalently $(\beta\!=\!-\sqrt{2},\gamma\!=\!0)$], at $\tilde{E}=0$.
$\beta=0$ is always an extremum, which is a local minimum (maximum) for 
$F>0$ ($F<0$), at $\tilde{E}=4h_0$.
Additional extremal points include saddle points at 
$[\beta_1\!>\!0,\gamma\!=\!0,\pi/3]$, $[\beta_2\!>\!0,\gamma\!=\!\pi/6]$ 
and a local maximum at $[\beta_3>0,\gamma=\pi/6]$. 
The saddle points, when exist, support 
a barrier separating the various minima, as seen in Fig.~\ref{fig1}. 
For large $N$, the normal modes involve 
$\beta$ and $\gamma$ vibrations about the 
respective deformed minima, with frequencies
\bsub
\ba
&&
\epsilon_{\beta 1}=\epsilon_{\beta 2} 
= \frac{8}{3}(h_0+ 2h_2)N^2 ~,
\\
&&\epsilon_{\gamma 1}=\epsilon_{\gamma 2} = 4\eta_3N^2 ~.
\ea
\label{d-modes}
\esub
For $h_0\!=\!0$, also $\beta\!=\!0$ becomes a global minimum, 
resulting in three degenerate minima corresponding to 
coexistence of prolate, oblate and 
spherical (S-P-O) shapes. $\hat{H}(h_0\!=\!0)$ satisfies 
Eq.~(\ref{HvanishPO}) and has also 
the following U(5) basis state
\ba
\hat{H}(h_0=0)\ket{N,n_d=\tau=L=0} = 0 ~,
\ea
as an eigenstate. Equivalently, it annihilates the intrinsic state 
of Eq.~(\ref{int-state}), with $\beta=0$. The additional normal modes 
involve quadrupole vibrations about the spherical minimum, with frequency
\ba
\epsilon = 4h_2N^2 ~.
\label{s-modes}
\ea

The members of the prolate and oblate ground-bands, 
Eq.~(\ref{HvanishPO}), 
are zero-energy eigenstates of $\hat{H}$ (\ref{Hint}), 
with good SU(3) and $\bsu3$ symmetry, respectively. 
The Hamiltonian is invariant under a change of sign 
of the $s$-bosons, hence commutes with the ${\cal R}_{s}$ operator 
mentioned above. 
Consequently, all non-degenerate eigenstates of $\hat{H}$ 
have well-defined $s$-parity. 
This implies vanishing quadrupole moments for an $E2$ operator 
which is odd under such sign change.
To overcome this difficulty, we introduce a small $s$-parity 
breaking term ${\textstyle\alpha\hat{\theta}_2 = 
\alpha[-\hat{C}_{2}[SU(3)] + 2\hat{N}(2\hat{N}+3)]}$, 
which contributes to $\tilde{E}(\beta,\gamma)$ a component 
${\textstyle\tilde{\alpha}(1+\beta^2)^{-2}[ 
(\beta^2\!-\!2)^2 \!+\! 2\beta^2(2 \!-\!2\sqrt{2}\beta\Gamma 
\!+\!\beta^2)]}$, 
with ${\textstyle\tilde{\alpha}=\alpha/(N-2)}$. 
The linear $\Gamma$-dependence distinguishes 
the two deformed minima and slightly lifts 
their degeneracy, as well as that of the normal modes~(\ref{d-modes}). 
Replacing ${\textstyle\hat{\theta}_2}$ 
by ${\textstyle\bar{\theta}_2 \!=\! 
-\hat{C}_{2}[\bsu3] + 2\hat{N}(2\hat{N}+3)}$, 
leads to similar effects but
interchanges the role of prolate and oblate bands. 
Identifying the collective part with $\hat{C}_2[{\rm SO(3)}]$, 
we arrive at the following complete Hamiltonian 
\ba
\hat{H}' &=& 
h_0\,P^{\dag}_0\hat{n}_sP_0 + h_2\,P^{\dag}_0\hat{n}_dP_0 
+\eta_3\,G^{\dag}_3\cdot\tilde{G}_3
+ \alpha\,\hat{\theta}_2 
+ \rho\,\hat{C}_2[\rm SO(3)] ~.\qquad
\label{Hprime}
\ea
\begin{figure}[t!]
  \centering
\includegraphics[width=11cm]{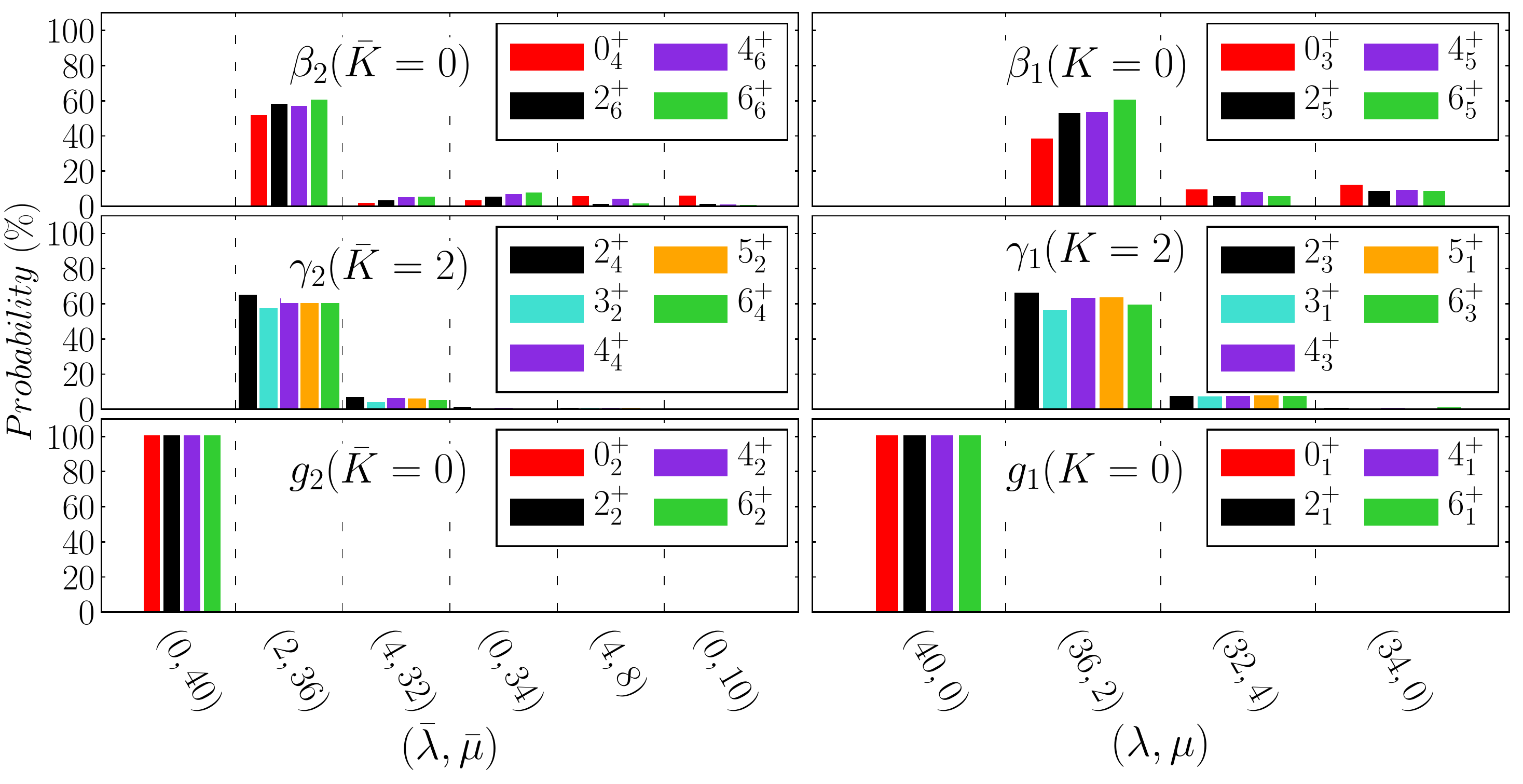}
\caption{
SU(3) $(\lambda,\mu)$- and $\bsu3$ $(\blam,\bmu)$-decompositions 
for members of the prolate ($g_1,\beta_1,\gamma_1$) 
and oblate ($g_2,\beta_2,\gamma_2$) bands, eigenstates of 
$\hat{H}'$ (\ref{Hprime}) with parameters as in Fig.~1(c), 
resulting in prolate-oblate (P-O) shape coexistence. 
Shown are probabilities larger than 5\%.
\label{fig2}}
\end{figure}
\begin{figure*}[t!]
  \centering
\includegraphics[width=11cm]{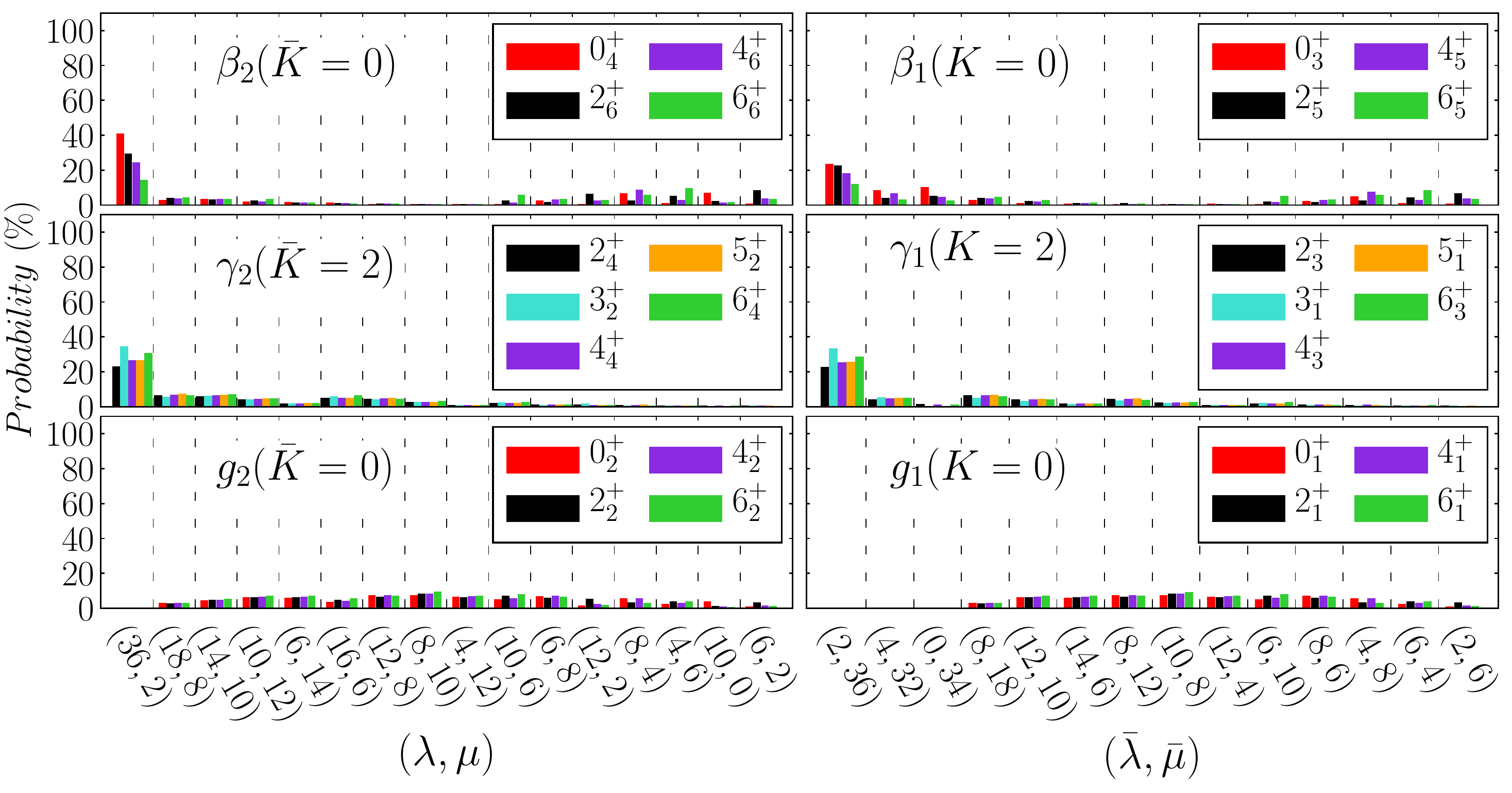}
  \caption{(Color online).
As in Fig.~\ref{fig2}, but now states of the 
prolate ($g_1,\beta_1,\gamma_1)$ bands 
are expanded in the $\bsu3$ basis, 
while states of the oblate ($g_2,\beta_2,\gamma_2)$ 
bands are expanded in the SU(3) basis. 
Shown are probabilities larger than 6\%.
\label{fig3}}
\end{figure*}
\begin{figure}[t]
  \centering
\includegraphics[width=11cm]{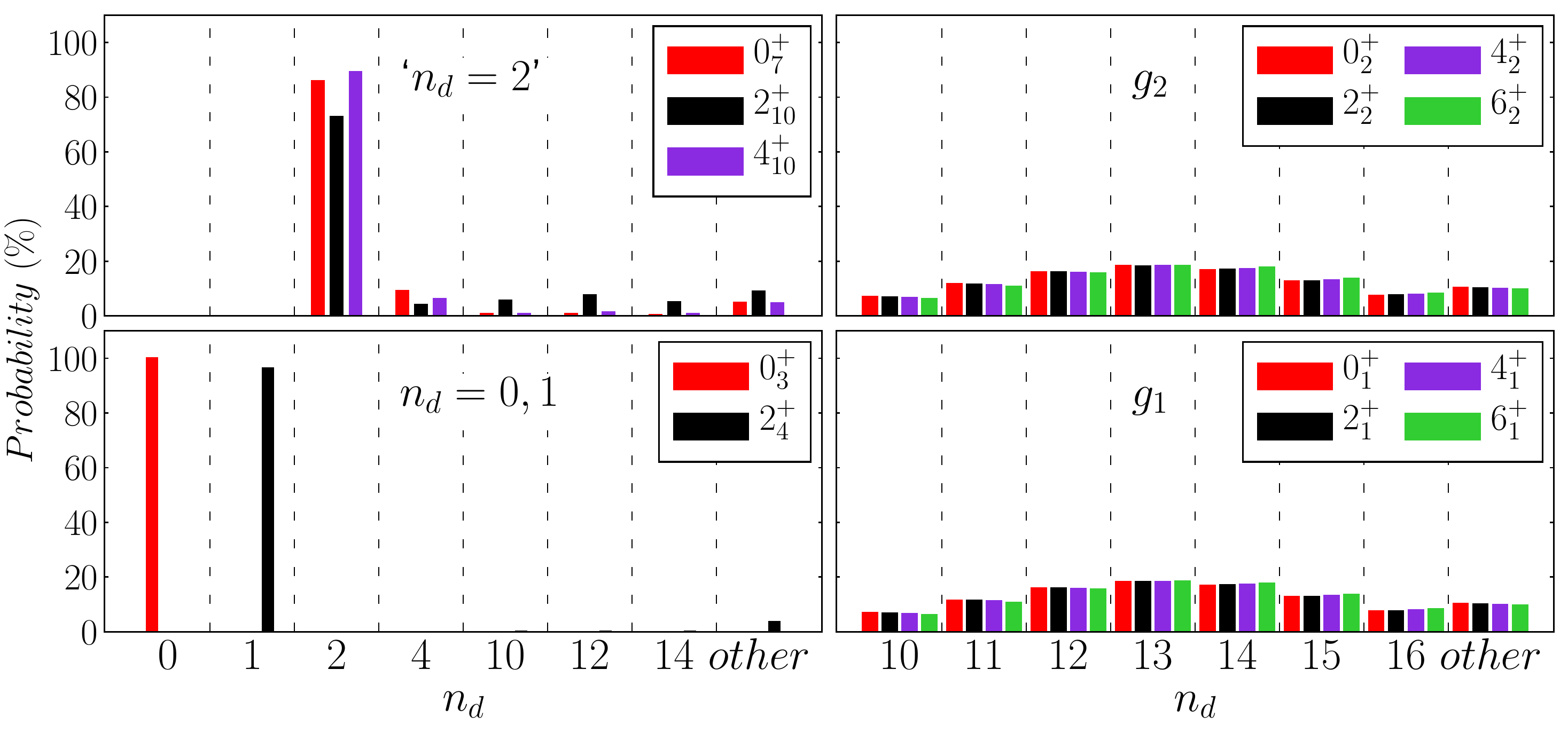}
\caption{
U(5) $n_d$-decomposition for spherical states (left panels)
and for members of the deformed prolate ($g_1$) and oblate ($g_2$) 
ground bands (right panels), 
eigenstates of 
$\hat{H}'$~(\ref{Hprime}) with parameters as in Fig.~1(f), 
resulting in spherical-prolate-oblate (S-P-O) shape coexistence.
The column `other' depicts a sum of probabilities, each less than 5\%. 
\label{fig4}}
\end{figure}
Figures 1(a)-1(b)-1(c) [1(d)-1(e)-1(f)] show 
$\tilde{E}(\beta,\gamma)$, $\tilde{E}(\beta,\gamma\!=\!0)$ 
and the bandhead spectrum of $\hat{H}'$~(\ref{Hprime}), 
with parameters ensuring degenerate P-O [S-P-O] minima. 
The prolate $g_1$-band 
remains solvable with energy $E_{g1}(L)=\rho L(L+1)$. 
The oblate $g_2$-band experiences a slight shift of
order ${\textstyle\tfrac{32}{9}\alpha N^2}$ and 
displays a rigid-rotor like spectrum. In the case of P-O coexistence, 
the SU(3) and $\bsu3$ decomposition in Fig.~\ref{fig2} demonstrates 
that these bands are pure DS basis states, with 
$(2N,0)$ and $(0,2N)$ character, respectively, 
while excited $\beta$ and $\gamma$ bands exhibit considerable mixing.
The critical-point Hamiltonian thus has a subset of states with good SU(3) 
symmetry, a subset of states with good $\bsu3$ symmetry and all other states 
are mixed with respect to both SU(3) and $\bsu3$. These are precisely the 
defining ingredients of SU(3)-PDS coexisting with $\bsu3$-PDS. 
The two persisting symmetries are incompatible, 
as is evident from Fig.~\ref{fig3}, 
where the same prolate ($g_1,\beta_1,\gamma_1$) bands are expanded in the 
$\bsu3$ basis, while the oblate ($g_2,\beta_2,\gamma_2$) bands are expanded 
in the SU(3) basis. All states, including the solvable ones, are 
seen to be strongly mixed and highly fragmented among many irreps. 

In the case of triple S-P-O coexistence, the prolate and oblate bands show 
similar behaviour. A new aspect is the 
simultaneous occurrence in the spectrum [Fig.~\ref{fig1}(f)] 
of spherical type of states, whose wave functions are dominated by 
a single $n_d$ component. 
As shown in Fig.~\ref{fig4}, 
the lowest spherical states have quantum numbers 
$(n_d\!=\!L\!=\!0)$ and $(n_d\!=\!1,L\!=\!2)$, 
hence coincide with pure U(5) basis states, while higher spherical states 
have a pronounced ($\sim$70\%) $n_d\!=\!2$ component. This structure 
should be contrasted with the U(5) decomposition of deformed states 
(belonging to the $g_1$ and $g_2$ bands) which, as shown in 
Fig.~\ref{fig4}, have a broad $n_d$-distribution.
The purity of selected sets of states 
with respect to SU(3), $\bsu3$ and U(5), in the presence of other 
mixed states, are the hallmarks of 
coexisting partial dynamical symmetries.

Since the wave functions for the members of the $g_1$ and $g_2$ bands 
are known, one can derive analytic expressions for their 
quadrupole moments and $E2$ transition rates. 
Considering the $E2$ operator 
$T(E2) = e_B\,\Pi^{(2)}$ with 
\ba
\Pi^{(2)} = d^{\dag}s + s^{\dag}\tilde{d} ~,
\label{Pi2}
\ea
the quadrupole moments 
are found to have equal magnitudes and opposite signs, 
\ba
Q_L &=& 
{\textstyle
\mp e_B\sqrt{\frac{16\pi}{40}}\frac{L}{2L+3}
\frac{4(2N-L)(2N+L+1)}{3(2N-1)}} ~,
\label{quadmom}
\ea
where the minus (plus) sign corresponds to the prolate-$g_1$ (oblate-$g_2$) 
band. The $B(E2)$ values for intraband ($g_1\to g_1$, $g_2\to g_2$) 
transitions, 
\ba
&&B(E2; g_i,\, L+2\to g_i,\,L) = 
\nonumber\\
&&
\quad
\;\;
{\textstyle
e_{B}^2\,\frac{3(L+1)(L+2)}{2(2L+3)(2L+5)}
\frac{(4N-1)^2(2N-L)(2N+L+3)}{18(2N-1)^2}} ~,
\qquad\qquad
\label{be2}
\ea
are the same. These properties are ensured by the fact that 
${\cal R}_sT(E2){\cal R}_s^{-1} = -T(E2)$. Interband 
$(g_2\leftrightarrow g_1)$ 
$E2$ transitions, are extremely weak. This follows from the fact that 
the $L$-states of the $g_1$ and $g_2$ bands exhaust, respectively, 
the $(2N,0)$ and $(0,2N)$ irrep of SU(3) and $\bsu3$. 
$T(E2)$ contains a $(2,2)$ tensor under both algebras, 
hence can connect the $(2N,0)$ irrep of $g_1$ only with the $(2N-4,2)$ 
component in $g_2$ which, as seen in Fig.~\ref{fig3}, 
is vanishingly small. The selection rule $g_1\nleftrightarrow g_2$ 
is valid also for a more general $E2$ operator, 
obtained by including in it the operators 
$Q^{(2)}$ or $\bar{Q}^{(2)}$, since the latter, as generators, 
cannot mix different irreps of SU(3) or $\bsu3$. 
By similar arguments, $E0$ transitions in-between 
the $g_1$ and $g_2$ bands are extremely weak, 
since the relevant operator, 
$T(E0)\propto\hat{n}_d$, is a combination of $(0,0)$ and $(2,2)$ 
tensors under both algebras. 
In contrast to $g_1$ and $g_2$, excited 
$\beta$ and $\gamma$ bands are mixed, hence are connected by 
$E2$ transitions to these ground bands. 

In the case of triple (S-P-O) coexistence, since $T(E2)$ 
obeys the selection rule $\Delta n_d\!=\!\pm 1$, the 
spherical states, $(n_d\!=\!L\!=\!0)$ and $(n_d\!=\!1,L\!=\!2)$,
have no quadrupole moment and the B($E2$) value for their 
connecting transition, obeys the U(5)-DS expression~\cite{ibm}
\ba
B(E2; n_d=1,L=2\to n_d=0,L=0) = e_{B}^2N ~.\quad
\label{be2nd}
\ea
These spherical states have very weak $E2$ transitions to the 
deformed ground bands, because they exhaust the $(n_d\!=\!0,1)$ irreps 
of U(5), and the $n_d\!=\!2$ component in the ($L\!=\!0,2,4$) states 
of the $g_1$ and $g_2$ bands is 
extremely small, of order $N^33^{-N}$, as seen in Fig.~\ref{fig4}. 
There are also no $E0$ transitions involving these spherical states, 
since $T(E0)$ is diagonal in $n_d$. The analytic expressions 
of Eqs.~(\ref{quadmom})-(\ref{be2nd}) are parameter-free predictions, 
except for a scale, and can be used 
to compare with measured values of these observables 
and to test the underlying SU(3), $\bsu3$ and U(5) partial symmetries.

\section{Departure from the Critical Point}

\begin{figure}[t!]
  \centering
\includegraphics[width=3.7cm]{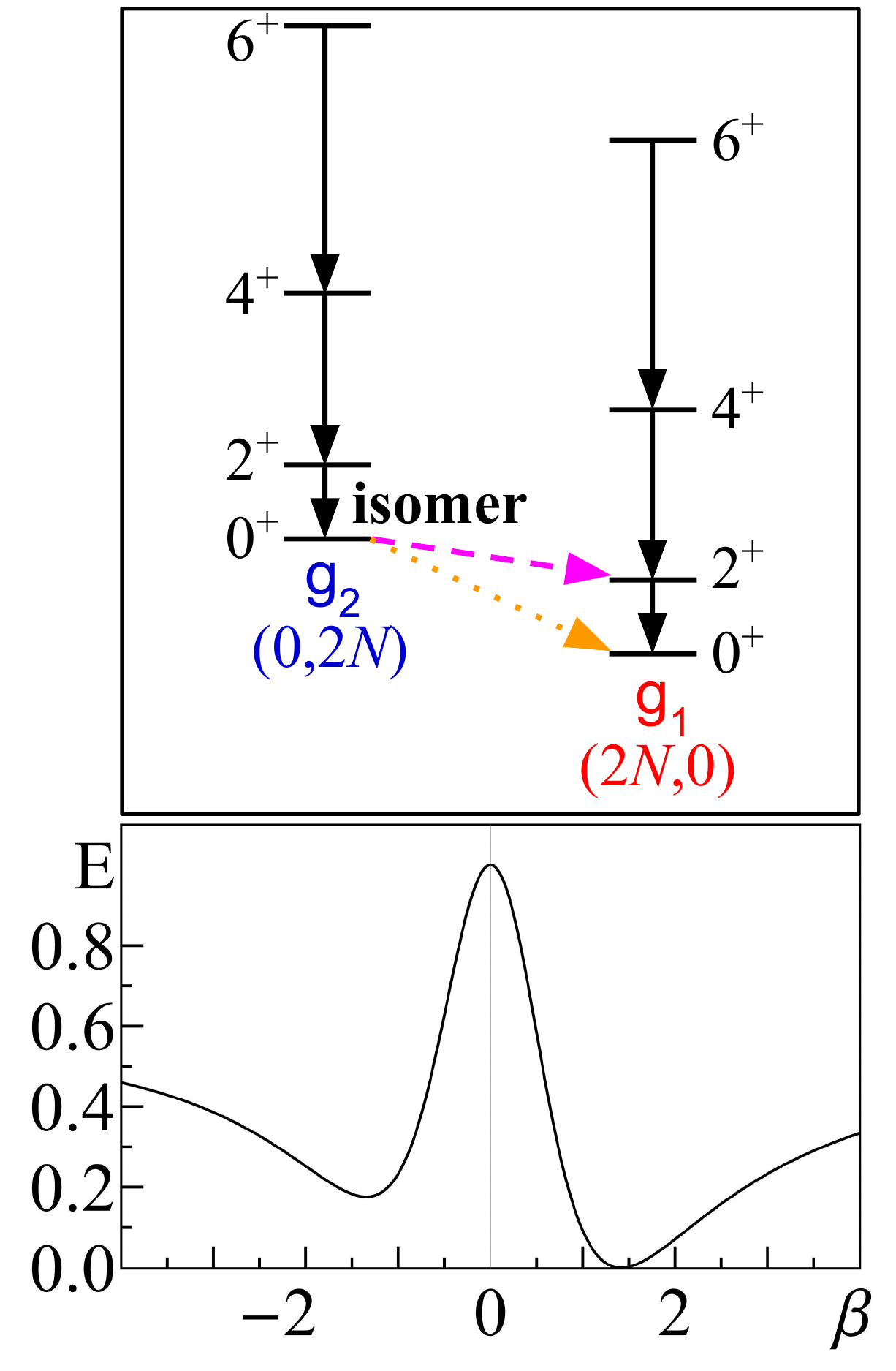}
\hspace{-0.15cm}
\includegraphics[width=3.7cm]{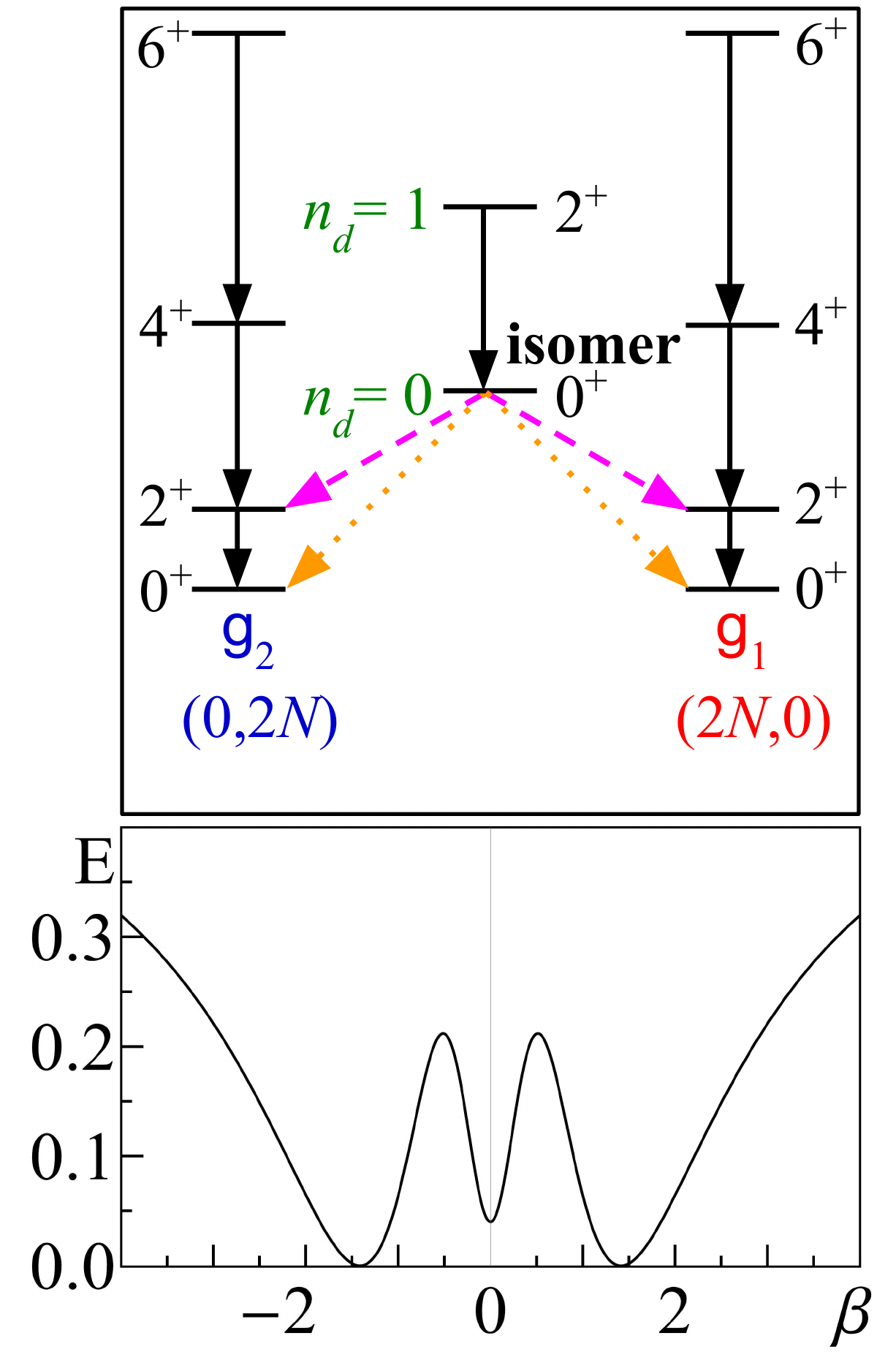}
\hspace{-0.15cm}
\includegraphics[width=3.7cm]{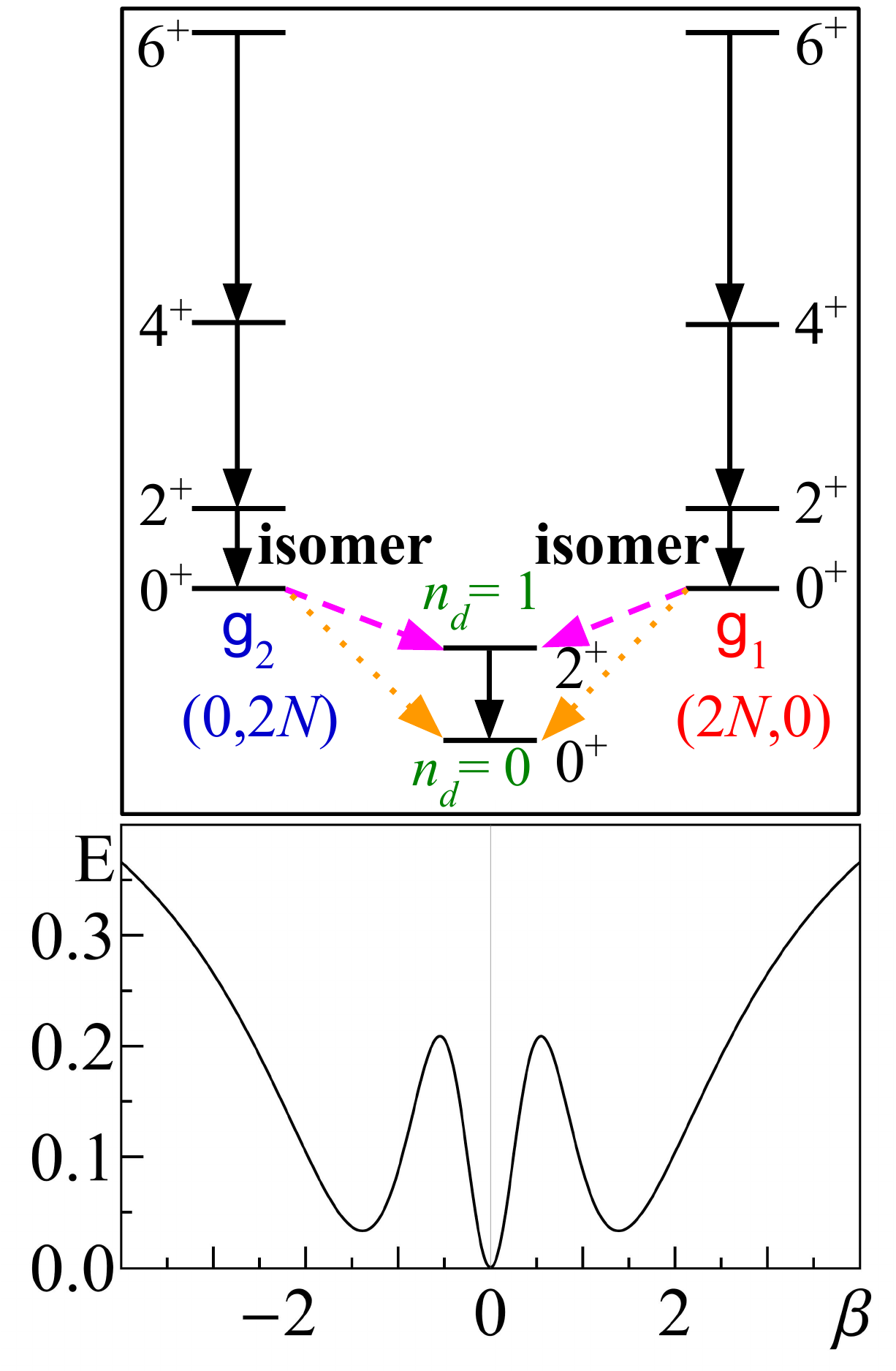}
\caption{\label{fig5}
Energy-surface sections and level schemes 
corresponding to departures from the critical point for 
$\hat{H}'$, Eq.~(\ref{Hprime}),
with $\eta_3\!=\!0.571,\,\rho\!=\!1$ and $N\!=\!20$. 
Left panels: P-O coexistence, oblate isomeric state 
($h_0\!=\!0.2,\,h_2\!=\!0.4,\,\alpha\!=\!0.9$). 
Middle panels: S-P-O coexistence, a spherical isomeric state 
($h_0\!=\!0.01,\,h_2\!=\!0.5,\,\alpha\!=\!0.018$). 
Right panels: S-P-O coexistence, deformed isomeric states 
($h_0\!=\!0,\,h_2\!=\!0.5,\,\alpha\!=\!0.018$ 
and an added $\epsilon\hat{n}_d$ term with $\epsilon\!=\!0.05$). 
Retarded $E2$ (dashes lines) and $E0$ (dotted lines) decays identify 
the isomeric states.}
\end{figure}

The above discussion has focused on the dynamics in the vicinity of the 
critical point where the multiple minima are near degenerate. 
The evolution of structure away from the critical point, can be studied by 
varying the coupling constants or by incorporating additional terms 
in $\hat{H}'$~(\ref{Hprime}). 
In case of P-O coexistence, taking larger values of $\alpha$, will leave 
the prolate $g_1$-band unchanged, but will shift the oblate $g_2$-band 
to higher energy of order $16\alpha N^2/9$. 
In case of triple S-P-O coexistence, 
if the spherical minimum is only local, 
one can use $\hat{H}'$~(\ref{Hprime}) 
with parameters satisfying $h_2\!>\!4h_0$, 
for which the spherical ground state $(n_d\!=\!L\!=\!0)$ 
experiences a shift of order $4h_0N^3$, but the deformed ground bands are
unchanged. Otherwise, if the deformed minima are only local, 
adding an $\epsilon\hat{n}_d$ term to $\hat{H}'(h_0\!=\!0)$ 
will leave the $n_d=0$ spherical ground state unchanged, but will shift 
the prolate and oblate bands to higher energy of order $2\epsilon N/3$. 
The resulting topology of the energy surfaces with such modifications 
are shown at the bottom row of Fig.~\ref{fig5}. If these departures 
from the critical points are small, the wave functions decomposition of 
Figs.~\ref{fig2}-\ref{fig4} remain intact and the analytic 
expressions for E2 observables 
and selection rules are still valid to a good approximation. 
In such scenarios, the lowest $L=0$ state of the non-yrast configuration 
will exhibit retarded $E2$ and $E0$ decays, hence will have the attributes 
of an isomer state, as depicted schematically on the top row 
of Fig.~\ref{fig5}.

\section*{Acknowledgments}

This work is supported by the Israel Science Foundation (Grant 586/16).

\end{document}